\documentclass[aps,prl,reprint,superscriptaddress,amsmath,amssymb,preprintnumbers,showpacs]{revtex4-1}
\usepackage[normalem]{ulem}

\usepackage{enumerate}
\usepackage{graphicx}
\usepackage{color}
\usepackage{float} 
\usepackage{xspace}
\usepackage{bm} 
\graphicspath{{./}}

\newcommand{\abs}[1]{\left| #1 \right|} 

\newcommand{\floqel}[1]{\xi_{E_\text{e}}}
\newcommand{\floqnuc}[1]{\xi_{E_\text{N}}}

\renewcommand{\vec}[1]{{\mathbf{\bm{#1}}}}

\newcommand{\tra}{{\text{tra}}}
\newcommand{\ter}{{\text{pol}}}
\newcommand{\anom}{{\text{anom}}}
\newcommand{\mix}{{\text{mix}}}
\newcommand{\semic}{{\text{SC}}}
\newcommand{\dres}[1]{\tilde{#1}}
\newcommand{\uv}[1]{\ensuremath{\hat{\mathbf{#1}}}} 

\begin{document}
\title{Characterizing Anomalous High-Harmonic Generation in Solids}
\author{Lun \surname{Yue}}
\email{lun\_yue@msn.com}
\affiliation{Department of Physics and Astronomy, Louisiana State University, Baton Rouge, Louisiana 70803-4001, USA}
\author{Mette B. \surname{Gaarde}}
\email{gaarde@phys.lsu.edu}
\affiliation{Department of Physics and Astronomy, Louisiana State University, Baton Rouge, Louisiana 70803-4001, USA}
\date{\today} 

\begin{abstract}
  Anomalous high-harmonic generation (HHG) arises in certain solids when irradiated by an intense laser field, as the result of a nonlinear perpendicular current akin to a Hall current. Here, we theoretically characterize the anomalous HHG mechanism, via development of an ab-initio methodology for strong-field laser-solid interaction that allows a rigorous decomposition of the total current. We identify two key characteristics of the anomalous harmonic yields: an overall increase with laser wavelength; and pronounced minima at certain intensities or wavelengths around which the emission time profiles drastically change. Such signatures can be exploited to disentangle the anomalous harmonics from the competing interband harmonics, and thus pave way for the experimental identification and time-domain control of pure anomalous harmonics.
\end{abstract}

\maketitle


High-harmonic generation (HHG) \cite{Ghimire2011NPhys, Ghimire2019review, Li2020review, Goulielmakis2022review}, i.e. the coherent production of high-energy photons by upconversion of the photon energy of the laser, represents a promising tool to probe the structure \cite{Vampa2015PRL, Luu2015Nature, Lakhotia2020Nature, Yu2022PRA, Uzan2020NPhoton}, topology \cite{Bauer2018PRL, Silva2019NPhoton, Chacon2020PRB, Juerss2020PRA, Schmid2021Nature, Bai2021NPhys, Baykusheva2021NanoLett}, electron-correlation \cite{Tancogne-Dejean2017PRL, Silva2018NPhoton, Murakami2021PRB, Jensen2021PRB, Hansen2022PRA}, and ultrafast electron dynamics \cite{Vampa2015Nature, Garg2016Nature, Neufeld2021PRL} of condensed-matter systems. Recently, it has been proposed that in crystalline systems with a non-vanishing Berry curvature, a laser field can induce a nonlinear anomalous current polarized perpendicular to the laser polarization and lead to the generation of non-perturbative anomalous high harmonics \cite{Liu2017NPhys, Luu2018NCommun, Silva2019NPhoton, Bai2021NPhys, Schmid2021Nature, Lv2021NCommun}. The Berry curvatures and anomalous carrier velocities are ubiquitous in modern condensed-matter theory and are e.g. responsible for various Hall effects \cite{Xiao2010review, Xiao2012PRL, Vanderbilt2018book}, and as such, the full understanding of anomalous current in the nonlinear regime is of utmost importance for both fundamental and application purposes.

The theoretical consensus of HHG consists of the intraband harmonics originating in the motion of the charge carriers in their respective bands, and the interband harmonics originating in the coherences between  bands. In calculations, the anomalous HHG mechanism has generally been treated as intraband \cite{Liu2017NPhys, Luu2018NCommun, Silva2019NPhoton, Chacon2020PRB, Wilhelm2021PRB, Schmid2021Nature, Lv2021NCommun}, and most frequently modeled semiclassically \cite{Liu2017NPhys, Luu2018NCommun, Silva2019NPhoton, Schmid2021Nature, Lv2021NCommun}, in terms of the perpendicular anomalous electron velocities governed by the Berry curvatures. In recent experimental works using linearly polarized drivers, the measured perpendicular-polarized harmonics have frequently been attributed entirely to the anomalous HHG mechanism \cite{Liu2017NPhys, Luu2018NCommun, Bai2021NPhys, Lv2021NCommun}. In principle, however, no symmetry rule \cite{Neufeld2019NCommun} prohibits the generation of perpendicular-polarized interband harmonics, as has been advocated in several recent works \cite{Kaneshima2018PRL, Yoshikawa2019NCommun, Liu2020NJP, Kobayashi2021US, Cao2021OE, Heide2022Optica}. A theoretical framework that consolidates these different interpretations of the perpendicular harmonics would be highly desired, and would facilitate the rigorous isolation of the anomalous HHG contribution through its characteristic signatures in the spectral and temporal domains. Sub-cycle emission characteristics has previously aided experiments in determining the underlying HHG mechanisms in quartz as intraband in origin using optical fields \cite{Garg2016Nature, Garg2018NPhoton}, and in ZnO as interband in origin using midinfrared fields \cite{Vampa2015PRL, Vampa2015Nature}.

In this Letter, we characterize the anomalous HHG mechanism, in terms of when it can dominate over other generation mechanisms, as well as how to understand and control its subcycle emission dynamics. For this purpose, we develop an ab-initio methodology for strong-field light-matter interaction that allows a rigorous decomposition of the total current into their structure-gauge-invariant constituents, without the explicit need for the construction of a periodic structure gauge \cite{Resta1994review, Vanderbilt2018book, Marzari2012review, Silva2019PRB, Yue2020PRA, Jiang2020PRB, Yue2022JOSABtutorial}. We consider monolayer MoS$_2$ as our specific target system for HHG, due to its potential technological applications \cite{Wang2012reviewTMD, Mak2016reviewTMD, Manzeli2017reviewTMD, Choi2017reviewTMD} and the absense of propagation effects. We identify two characteristic signatures of the anomalous harmonics: (i) the yield increases with laser wavelength and dominates over interband harmonics below approximately 1 eV; (ii) exhibit characteristic minima in the yield at certain laser wavelengths or intensities around which the sub-cycle time structure changes. We characterize the ab-initio results using a semiclassical model for the anomalous current, whereby signature (ii) is traced to a phase-jump in the harmonic spectral amplitudes. The framework presented here adds to the fundamental understanding of HHG in solids, and can be a guide toward experimental identification and control of purely anomalous harmonics.


Our ab-initio methodology starts with the calculation of the monolayer MoS$_2$ band structure $E_n^{\vec{k}}$ and momentum coupling matrix elements $\vec{p}_{mn}^{\vec{k}}$, employing density functional theory \cite{Giannozzi2009codeQE, Giannozzi2017codeQE} with ONCV pseudopotentials \cite{Hamann2013PRB, Schlipf2015CPC} and PBE functionals \cite{Perdew1996PRL}. 
Atomic units is used throughout this work unless indicated otherwise.
The Brillouin zone is sampled with $N=8100$ points in a Monkhorst-Pack mesh. For the dynamics, we solve the time-dependent equations for the density matrix elements $\rho_{mn}^{\vec{k}}(t)$ in the velocity gauge \cite{Yue2020PRA, Yue2022JOSABtutorial, suppmat2022_mos2_anom}, taking into account 90 bands. At every 5th step of the time propagation, we transform into an adiabatic basis \cite{Yue2022JOSABtutorial} to include a dephasing time $T_2=10$ fs, and decompose the total current into four structure-gauge-invariant terms, $\vec{j}(t) = \vec{j}^\tra(t) + \vec{j}^\ter(t) + \vec{j}^\anom(t) + \vec{j}^\mix(t)$, where
\begin{subequations}
  \label{eq:meth_1}
  \begin{align}
    \vec{j}^\tra(t) &= - N^{-1}\sum_{n\vec{k}} \tilde{\vec{p}}_{nn}^{\vec{k}}(t) \tilde{\rho}_{nn}^{\vec{k}}(t), \label{eq:meth_1a} \\
    \vec{j}^\ter(t) &= - N^{-1}\partial_t \sum_{m\ne n, \vec{k}} \dres{\vec{d}}_{mn}^{\vec{k}}(t)
                    \tilde{\rho}_{nm}^{\vec{k}}(t), \label{eq:meth_1b} \\
    \vec{j}^\anom(t) &= - N^{-1} \sum_{n\vec{k}} \left[
                       \vec{F}(t) \times
                       \dres{\vec{\Omega}}_n^{\vec{k}}(t)
                     \right] \tilde{\rho}_{nn}^{\vec{k}}(t), \label{eq:meth_1c} \\
    \vec{j}^\mix(t) 
                    &= - N^{-1} \sum_{\mu=x,y} F_\mu(t) \sum_{m\ne n, \vec{k}}
                      \big[ \dres{d}_{\mu, mn}^{\vec{k}}(t) \big]_{;\vec{k}}
                      \tilde{\rho}_{nm}^{\vec{k}}(t) \label{eq:meth_1d},
  \end{align}
\end{subequations}
with $\vec{j}^\tra(t)$ the intraband current, $\vec{j}^\ter(t)$ the interband (polarization) current, $\vec{j}^\anom(t)$ the anomalous current, and $\vec{j}^\mix(t)$ the mixture current originating in the coupling between the intraband and interband position operators \cite{Blount1962book, Aversa1995PRB, Wilhelm2021PRB, Yue2022JOSABtutorial}. The tildes in Eq.~\eqref{eq:meth_1} indicate quantities evaluated in the adiabatic basis. After calculation of the current decomposition at time $t$, we transform back into the Bloch basis and continue with the time-dependent propagation of $\rho_{mn}^{\vec{k}}(t)$. We evaluate the dipole matrix elements $\dres{\vec{d}}_{mn}^{\vec{k}}(t)$, generalized gradients $\bigl[\dres{d}_{\mu, mn}^{\vec{k}}(t)\big]_{;\vec{k}}$ \cite{Aversa1995PRB, Sipe2000PRB}, and Berry curvatures $\dres{\vec{\Omega}}_n^{\vec{k}}(t)$ entirely in terms of $\dres{E}_n^{\vec{k}}(t)$ and $\dres{\vec{p}}_{mn}^{\vec{k}}(t)$ by using relevant sum-rules \cite{Gradhand2012review, Aversa1995PRB, Sipe2000PRB}. The key advantage of this methodology is that it does not require the construction of a periodic structure gauge for the Bloch states, in contrast to previous methods that decompose the current \cite{Aversa1995PRB, Wilhelm2021PRB}, but instead requires more bands for convergence (see Supplemental Material (SM) \cite{suppmat2022_mos2_anom}). The electric field, $\vec{F}(t)=-\partial_t \vec{A}(t)$, is linearly polarized in the plane of the monolayer, and the vector potential $\vec{A}(t)$ has a total duration of 12 optical cycles and a $\cos^2$-envelope. Since $\vec{j}^\anom(t) \perp \vec{F}(t)$, we focus our attention on the harmonic spectra polarized perpendicular to the laser field, $S(\omega) =\omega^2 \abs{\vec{j}(\omega)\cdot \uv{e}_{\perp}}^2$, with $\vec{j}(\omega)=\int_{-\infty}^\infty dt e^{i\omega t}\vec{j}(t)$.

We highlight that we do not label the anomalous current as an intraband contribution: even though $\vec{j}^\anom(t)$ in Eq.~\eqref{eq:meth_1c} does not depend on the time-dependent coherences, the Berry curvature for the $n$th band arises from the residual coupling between the other bands \cite{Xiao2010review, Gradhand2012review, suppmat2022_mos2_anom}. In addition, it can be easily shown that the oft-used expression for the ``interband'' current $N^{-1}\sum_{m\neq n, \vec{k}} \tilde{\vec{p}}_{mn}^{\vec{k}}(t) \tilde{\rho}_{nm}^{\vec{k}}(t)$ equals $\vec{j}^\ter(t)+\vec{j}^\anom(t)+\vec{j}^\mix(t)$. In this sense, the experimentally measured perpendicular harmonics from the literature interpreted as ``interband'' \cite{Kaneshima2018PRL, Yoshikawa2019NCommun, Liu2020NJP, Kobayashi2021US, Cao2021OE, Heide2022Optica} likely already contains the anomalous contribution, while in this work interband current refers to $\vec{j}^\ter$(t) of Eq.~\eqref{eq:meth_1b} \cite{Vampa2014PRL}.  Considering $\vec{j}^\anom(t)$ as a stand-alone contribution to the total current, facilitates a link between the weak-field condensed matter community, where Berry curvatures and anomalous velocities are ubiquitous, and the strong-field community. As we will see, the anomalous high harmonics exhibit unique characteristics separating them from the other mechanisms for HHG.


\begin{figure}[H]
  \centering
   \includegraphics[width=0.48\textwidth, clip, trim=0 0cm 0 0cm]{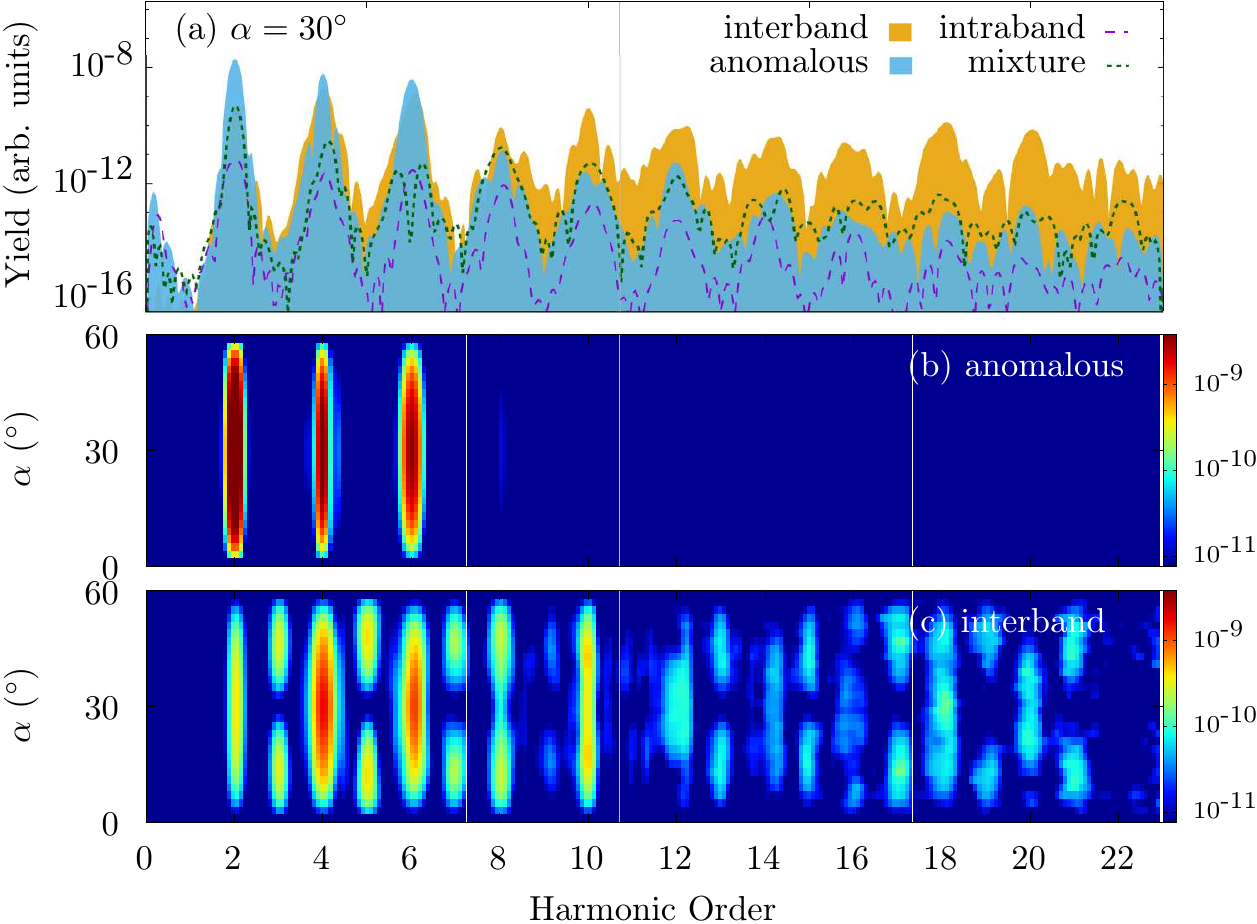}
  \caption{Spectra for perpendicular-polarized harmonics in monolayer MoS$_2$ for a 7.4 $\mu$m driver.
    (a) Decomposed spectra in terms of different contributions to the current for $\alpha=30^\circ$. The (b) anomalous and (c) interband harmonics versus $\alpha$. The gray vertical lines mark the gap energy.}
  \label{fig:anglescan}
\end{figure}

Figure \ref{fig:anglescan}(a) shows the decomposed perpendicular-polarized harmonic spectrum for a field with carrier wavelength 7.4 $\mu$m, intensity 50 GW/cm$^2$ and $\alpha=30^\circ$, where $\alpha$ denotes the angle between the field-polarization direction and the MoS$_2$ mirror plane. Only even-order harmonics are observed, with the interband and anomalous contributions clearly competing: the interband harmonics dominate at higher orders, even well below the band gap energy at around harmonic 8 (H8) and H10; while the anomalous harmonics dominate for the lowest orders at H2-H6. The intraband and mixture currents contribute with yields that are order(s) of magnitude lower.

Figures~\ref{fig:anglescan}(b) and \ref{fig:anglescan}(c) show respectively the anomalous and interband HHG spectra versus $\alpha$. Our results conform with the dynamical-symmetry selection rules for HHG \cite{Neufeld2019NCommun}, where the $D_{3h}$ spatial point group of monolayer MoS$_2$ together with the in-plane linearly-polarized field prohibit the emission of all perpendicular harmonics at $\alpha=n60^\circ$ and odd-order harmonics at $\alpha=30^\circ+n60^\circ$, with $n$ an integer \cite{Neufeld2019NCommun, Liu2017NPhys, Baykusheva2021PRA, Yue2022arxiv}. 
The anomalous harmonics in Fig.~\ref{fig:anglescan}(b) only permit even orders for all $\alpha$, since time-reversal symmetry dictates $\vec{\Omega}_m^{-\vec{k}}=-\vec{\Omega}_m^{\vec{k}}$ which leads to $\vec{j}^\anom(t+\frac{T}{2})=\vec{j}^\anom(t)$, with $T$ the optical period. In Fig.~\ref{fig:anglescan}(c), no symmetry rule prohibits the emission of interband harmonics between $\alpha=0^\circ$ and $\alpha=30^\circ$, and both even and odd orders are observed.


\begin{figure}[H]
  \centering
  \includegraphics[width=0.48\textwidth, clip, trim=0 0cm 0 0cm]{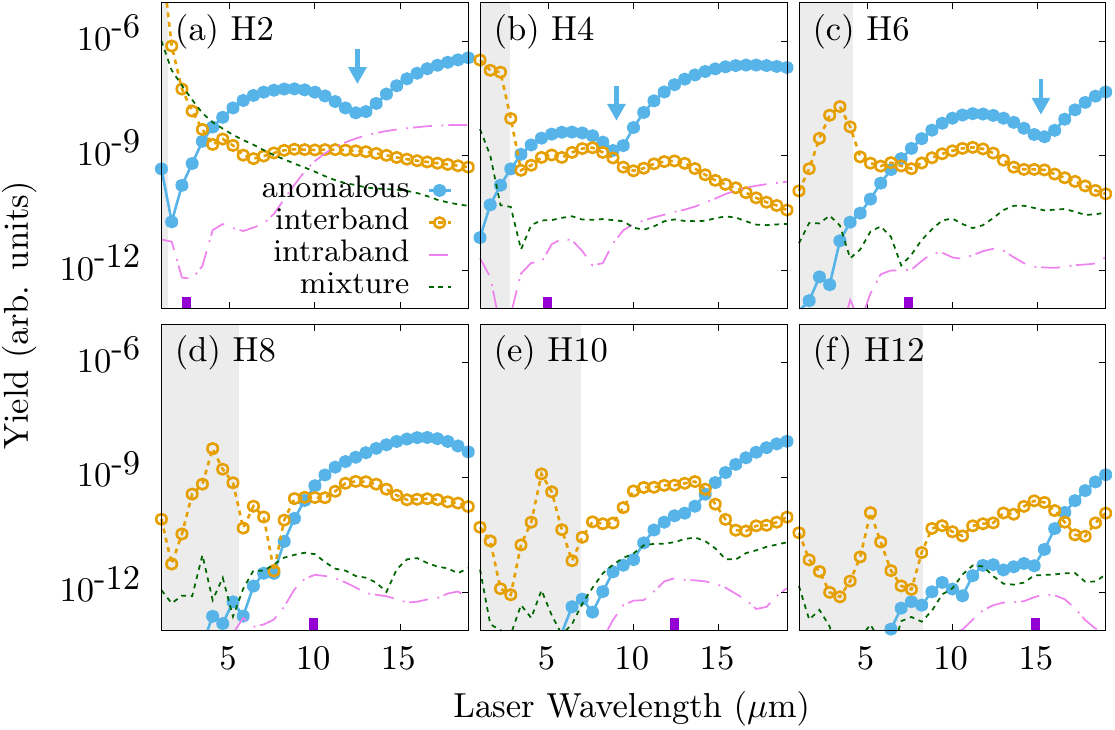}
  \caption{Decomposed harmonic yields for different laser wavelengths and harmonic orders at $\alpha=30^\circ$. The arrows in (a)-(c) mark pronounced minima in the anomalous harmonic yield. In each panel, the gray shaded area indicate the wavelength region where the harmonic energy is greater than the band gap energy, and the purple ticks on the wavelength axis mark the harmonic energy 1 eV.}
  \label{fig:wlscan}
\end{figure}

To further characterize the anomalous HHG mechanism and its competition with the other generation mechanisms, we show in Fig.~\ref{fig:wlscan} the decomposed harmonic yields for H2-H12 versus laser wavelength, fixing the laser intensity at 50~GW/cm$^2$ and $\alpha=30^\circ$. We focus on the anomalous and interband contributions, which are seen to dominate over the intraband and mixture contributions for almost all wavelengths. At shorter laser wavelengths, in the gray shaded regions of Fig.~\ref{fig:wlscan} where the harmonic energy is greater than the gap energy of 1.79 eV, the interband yield dominates over the anomalous yield for all harmonic orders. With increasing wavelength, the anomalous harmonic yield overall increases, and begins to dominate over the interband yield at around a harmonic energy of $E_{h}=1$ eV (marked by purple ticks on the wavelength axis). At the longest wavelength in the figure, 19 $\mu$m, the anomalous harmonics dominate completely -- e.g. the anomalous H6 yield at 19 $\mu$m is around 3 orders of magnitudes greater than the interband yield. We judge the transition energy $E_h$ to be material dependent and governed by the particular band structure and coupling matrix elements of the system, but nonetheless expect it to be generally smaller than the gap energy, as above-band-gap harmonics generally proceeds via interband coherences \cite{Vampa2014PRL, Wu2015PRA}. Figure~\ref{fig:wlscan} clears up the recent contention on the origin of the perpendicular harmonics in MoS$_2$ in terms of either anomalous \cite{Liu2017NPhys, Liu2020NJP, Lee2021Symmetry} or interband harmonics \cite{Yoshikawa2019NCommun, Liu2020NJP, Kobayashi2021US, Cao2021OE, Lee2021Symmetry, Heide2022Optica}: these two harmonic contributions dominate in different laser-parameter and harmonic-energy regimes.

In Figs.~\ref{fig:wlscan}(a)-\ref{fig:wlscan}(c), the anomalous harmonic yield for H2, H4 and H6 exhibit prominent minima marked by blue arrows at 12.5~$\mu$m, 9~$\mu$m and 15.2~$\mu$m, respectively. We performed additional HHG simulations using a fixed laser wavelength of 7.4 $\mu$m and varying laser intensities, and observed similar pronounced minima at specific intensities (see SM \cite{suppmat2022_mos2_anom}). These minima represent a striking characteristic of the anomalous harmonics, which we now characterize employing a semiclassical model.


\begin{figure}[H]
  \centering
  \includegraphics[width=0.48\textwidth, clip, trim=0 0cm 0 0cm]{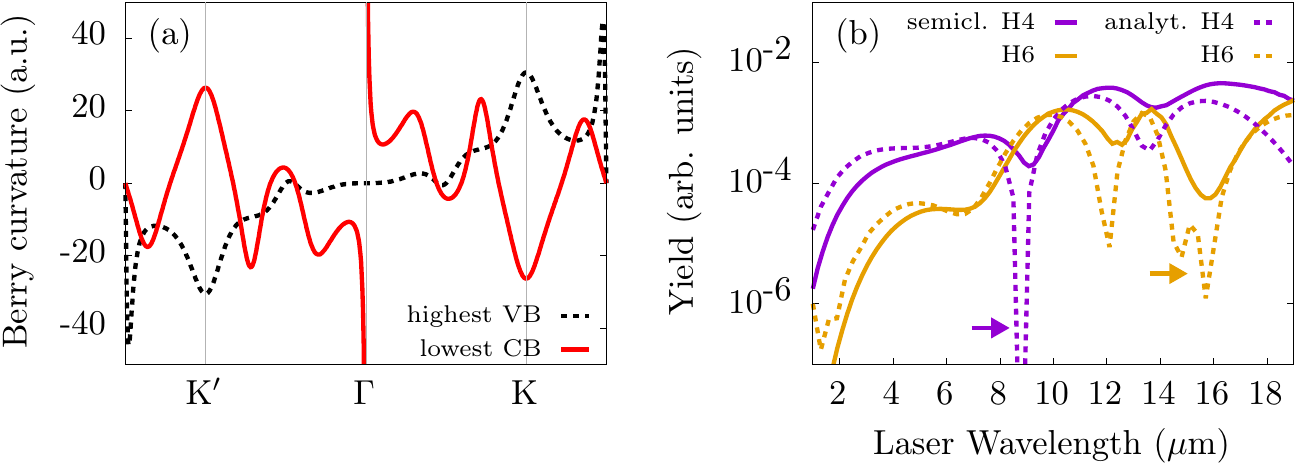}
  \caption{(a) Berry curvature along a line in reciprocal space for the highest VB and the lowest CB. (b) Semiclassical numerical and analytical results (see text) for the anomalous H4 and H6 yields versus wavelength.}
  \label{fig:berrycurv}
\end{figure}

We write the semiclassical anomalous current as $\vec{j}^\semic(t)\propto \vec{F}(t) \times \sum_{m}  \vec{\Omega}_m^{\vec{k}}\rho_{mm}^{\vec{k}}(t)$. Since tunnel excitation is most probable at the band gap \cite{Keldysh1964}, we only sum over the highest VB (VB1) and the lowest CB (CB1), and approximate the carriers as point particles at the $K$ and $K'$ symmetry points, leading to the density approximation $\rho_{mm}^{\vec{k}}=\left(\delta_{\vec{k}-\vec{A}(t),K} + \delta_{\vec{k}-\vec{A}(t),K'}\right)$  \cite{Liu2017NPhys}. For $\alpha=30^\circ$, we show in Fig.~\ref{fig:berrycurv}(a) the $z$-components of the Berry curvatures for VB1 and CB1 along the line in reciprocal space going through $K'$, $\Gamma$ and $K$. The semiclassical results, using the same laser fields as in the ab-initio calculations of Fig.~\ref{fig:wlscan}, are plotted in Fig.~\ref{fig:berrycurv}(b) for H4 and H6 in solid lines. Clear minima are observed, with the most pronounced minimum seen at $\sim$9~$\mu$m for H4 and at $\sim$15.7~$\mu$m for H6, consistent with our ab-initio results in Fig.~\ref{fig:wlscan}. To gain further insight, we assume a monochromatic vector potential $\vec{A}(t)=F_0/\omega_0\sin(\omega_0 t)\uv{e}$, which leads to the analytical expression for the current, $\vec{j}^\semic(t)\propto \sum_{s=1}^\infty C_{2s}\sin(2s\omega_0 t)$, with
\begin{equation}
  \label{eq:semicl_1}
  C_{2s}=\frac{s\omega_0}{a}
  \sum_{n}^{\left\{\text{VB1},\text{CB1}\right\}}
  \sum_{q=1}^{\infty}
  \frac{\Omega_{n, q}}{q} \cos\left(\frac{2q\pi}{3}\right)  J_{2s}\left(\frac{qa F_0}{2\omega_0}\right)
\end{equation}
the spectral amplitudes, $\Omega_{n,q}$ the Fourier coefficients of the Berry curvature for the $n$th band, $J_{2s}$ the $(2s)$th-order Bessel function, and $a$ the lattice constant. The yield versus wavelength for the $(2s)$th anomalous harmonic, $S_{2s} \propto s^2\omega_0^2\left|C_{2s}\right|^2$, is plotted in Fig.~\ref{fig:berrycurv}(b) for $2s=4$ and $2s=6$ in dashed lines. The pronounced minima marked by the arrows hence correspond to nodes of $C_{2s}$ seen as a function of wavelength. We see that the expression for $C_{2s}$ in Eq.~\eqref{eq:semicl_1} involves both the nonlinearities of the Bessel functions and the Berry curvature expansion coefficients $\Omega_{n,q}$. We have checked that the $q$-sum in Eq.~\eqref{eq:semicl_1} is not dominated by a single term, but rather a superposition of terms. In addition, the pronounced minima in the yield-versus-laser-intensity plots from the ab-initio calculations are also explained in terms of the nodes of $C_{2s}$ (considered as a function of the laser intensity) \cite{suppmat2022_mos2_anom}. The nodes in $C_{2s}$ is another characterization tool for the anomalous harmonics, as the presence of such minima in a wavelength or intensity scan is an indication of the dominance of the anomalous HHG mechanism.


\begin{figure}[H]
  \centering
  \includegraphics[width=0.48\textwidth, clip, trim=0 0cm 0 0cm]{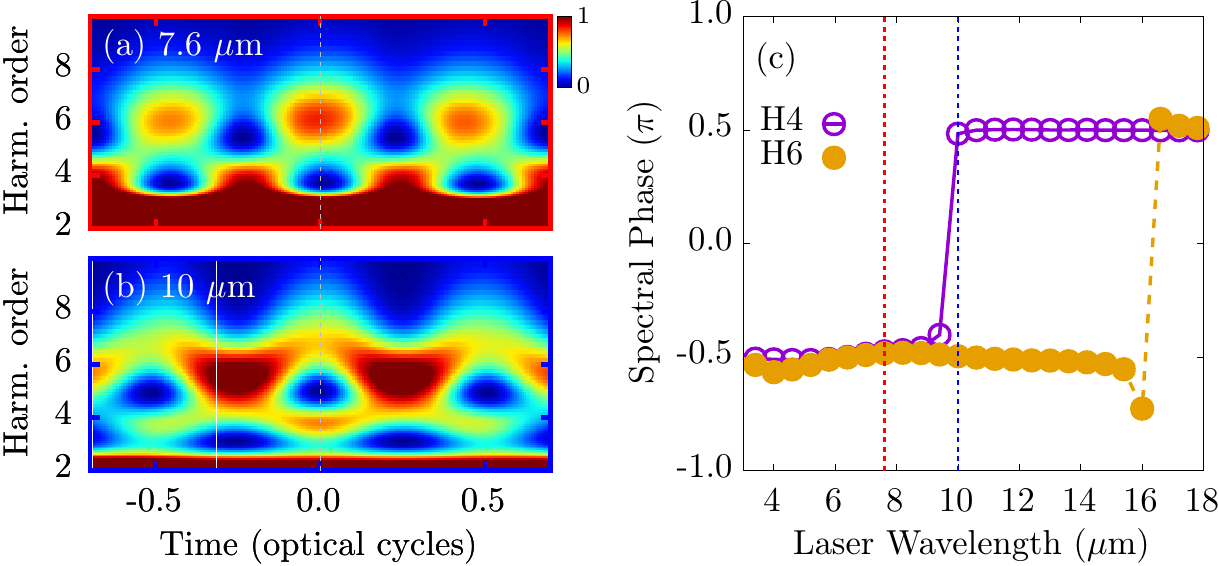}
  \caption{Time-frequency emission profiles for the anomalous harmonics extracted from the ab-initio simulations for laser wavelength (a) 7.6 $\mu$m and (b) 10 $\mu$m. The colorbar axis is in linear scale and arbitrary units, and the dashed vertical line is to guide the eyes. (c) Spectral phases for the anomalous H4 and H6 extracted from the simulations, with the wavelengths used in (a) and (b) marked by vertical lines.}
  \label{fig:cwt}
\end{figure}

Finally, we consider the emission-time profiles and the spectral phases of the solid-state harmonics, which encode pertinent information about the underlying harmonic emission mechanisms \cite{Vampa2015PRL, Vampa2015Nature, Garg2016Nature, Garg2018NPhoton}.
In Figs.~\ref{fig:cwt}(a) and \ref{fig:cwt}(b), we show the time-frequency profiles for the anomalous harmonic emissions obtained from the ab-initio calculations, at laser wavelengths corresponding to before and after the H4 minimum at $\lambda_{min}\equiv 9$ $\mu$m [Fig.~\ref{fig:wlscan}(b)], respectively. For the 7.6 $\mu$m case in Fig.~\ref{fig:cwt}(a), the spectral yields around H6 is emitted at $t=T/2$, corresponding to the electric field extrema, while H4 is emitted at $t=(n+\frac{1}{2})T/2$ at the zeros of the field. For the longer laser wavelength 10 $\mu$m $>\lambda_{min}$ in Fig.~\ref{fig:cwt}(b), the emission time of H4 and H6 are shifted by a quarter optical cycle.

These shifts of the anomalous harmonic emission times can be understood in terms of the semiclassical analytical model for the anomalous current. As we have seen, the yield minima of the ($2s$)th harmonic is explained by the nodes of spectral amplitudes $C_{2s}$. When going through such a node, the real-valued spectral amplitude $C_{2s}$ changes sign, corresponding to a $\pi$-phase shift, which in turn will change the emission-time profiles. We confirm our interpretation by extracting the spectral phases of the anomalous harmonics from our ab-initio simulations, $\varphi^\anom(\omega)=\arg[\vec{j}^\anom(\omega)\cdot \uv{e}_{\perp}]$, and present the results in Fig.~\ref{fig:cwt}(c). The spectral phases for H4 and H6 for the laser wavelength range 3 $\mu$m to 9 $\mu$m are in-phase with value $-\pi/2$. A sharp $\pi$-phase shift occurs for H4 at the wavelength of the yield minimum $\sim9$ $\mu$m, such that H4 and H6 becomes out-of-phase from 10 $\mu$m to 16 $\mu$m. This change of the spectral phase from being in-phase to out-of-phase explains the time-shift of the emissions in Figs.~\ref{fig:cwt}(a) and \ref{fig:cwt}(b). At $\sim 16$ $\mu$m, the wavelength corresponding to the yield minimum of H6 in Fig.~\ref{fig:wlscan}(c), the spectral phase of H6 shifts by $\pi$. Additional calculations confirm that the phase-shift of the anomalous HHG also occurs around the minima of a laser-intensity scan, again validating the robustness of our ab-initio results and our semiclassical interpretations. In addition to the emission-time shifts and the sharp spectral-phase shifts being unique characteristics of the anomalous HHG mechanism, they also suggest the potential control of anomalous harmonics by taking advantage of the sign change around the nodes of $C_s$.


 In conclusion, we have uncovered distinct characteristics of the anomalous HHG mechanism in MoS$_2$. First, we found that the anomalous harmonic yield increases with laser wavelength, and dominates over the interband yield below a certain energy threshold $E_h$. Second, we identified that the anomalous harmonic yield exhibits minima at certain field parameters, around which the sub-cycle emission-time profile changes. We used a semiclassical model to trace these minima to nodes in the spectral amplitudes of the anomalous harmonics.

 We expect our findings to be general and directly applicable to other materials that have finite Berry curvatures. Indeed, neither the anomalous nor the interband perpendicular harmonics are prohibited by symmetry, and will generally be competing. The presence of minima in the anomalous harmonic yields are also of a general nature, since the derivation of the semiclassical spectral amplitude in Eq.~\eqref{eq:semicl_1} will always contain Bessel functions. With our rigorous theoretical methodology, we have shed light on a recent subject of intense debate on whether the perpendicular-polarized harmonics originate from the anomalous current \cite{Liu2017NPhys, Luu2018NCommun, Bai2021NPhys, Lv2021NCommun} or the interband current \cite{Kaneshima2018PRL, Yoshikawa2019NCommun, Kobayashi2021US, Cao2021OE, Heide2022Optica}. Our results open up the possibility for the experimental identification and time-domain control of pure anomalous harmonics, which we believe are accessible with current technologies. Our findings will also be valuable for the reconstruction of Berry curvatures using HHG \cite{Liu2017NPhys, Luu2018NCommun, Lv2021NCommun}, as such schemes requires the measurement of pure anomalous harmonics.

 Finally, all our findings are facilitated by the development of an general ab-initio methodology that allows the rigorous decomposition of the currents without the construction of a globally periodic structure gauge, which can directly be applied to other complex systems. We expect our approach to be especially useful for ab-initio simulations of strong-field interactions in topological insulators such as Chern insulators, where a nonzero Chern number prevents the construction of a globally periodic structure gauge \cite{Thonhauser2006PRB, Marzari2012review}.

\acknowledgements
L. Y. and M. B. G. acknowledge support from the National Science Foundation, under Grant PHY-2110317, and Air Force Office of Scientific Research, under Grant No. FA9550-16-1-0013. L. Y. acknowledges Shicheng Jiang for helpful insights at the initial stage of the project, and thanks Michael Zuerch, Yuki Kobayashi and Ofer Neufeld for useful discussions.


%

\end{document}